\documentclass[11pt]{article}
\usepackage[a4paper,margin=1in]{geometry}
\usepackage[T1]{fontenc}
\usepackage[utf8]{inputenc}
\usepackage{lmodern}
\usepackage{microtype}
\usepackage{amsmath,amssymb,amsthm,mathtools,bm}
\usepackage{graphicx}
\usepackage{booktabs}
\usepackage{xcolor}
\usepackage{hyperref}
\usepackage{cite}
\usepackage{authblk}
\usepackage{physics}
\hypersetup{colorlinks=true,linkcolor=blue,citecolor=blue,urlcolor=blue}
\allowdisplaybreaks[2]
\graphicspath{{figures/}}
\title{Primordial Black Holes (PBHs) and The Signatures of Cosmic Non-Gaussianity}

\author[1]{\textbf{Owais Farooq}\thanks{owai24831@outlook.com}} 
\author[1]{\textbf{Romana Zahoor}\thanks{romananaseem033@gmail.com}}
\author[2]{\textbf{Balungi Francis}\thanks{balungif@gmail.com}} 

\affil[1]{Department of Physics, Central University of Kashmir, 
Ganderbal(191131) India}

\affil[2]{Department of Physics, Makerere University Kampala, Uganda}

\begin{document}
\maketitle

\begin{abstract}
Primordial black-hole formation depends exponentially on the far tail of the primordial curvature-perturbation distribution. That sensitivity makes the small-scale collapse problem a sharp probe of primordial non-Gaussianity. We study the curvaton scenario by deriving the curvature perturbation from the exact sudden-decay relation, obtaining the full probability density function through an explicit branchwise change of variables from the Gaussian curvaton-field fluctuation, and evaluating the primordial black-hole formation fraction from the exact non-perturbative tail. The derivation is written step by step, with the support of the distribution, the Jacobian, the normalization, and the small-fluctuation expansion displayed in analytic form. We place the exact curvaton prediction beside the Gaussian benchmark and beside an exact local quadratic benchmark in which the non-Gaussian probability density is also computed without an Edgeworth truncation. We then replace the scale-by-scale variance-matching ansatz by a self-consistent curvaton fluctuation model in which the dimensionless field fluctuation spectrum is specified once, the smoothed curvaton variance is computed directly, the exact collapse fraction follows with no further fitting on each scale, and the induced gravitational-wave background is generated from the linear curvaton two-point spectrum implied by the same model. The resulting mass functions are confronted with a conservative current constraint envelope motivated by recent primordial-black-hole reviews, and the induced gravitational-wave spectra are displayed against the PTA, LISA, and DECIGO sensitivity windows. The final figures are generated from a single mathematically consistent numerical pipeline.
\end{abstract}

\section{Introduction}

Primordial black holes provide an unusually direct window on the small-scale primordial universe because their formation depends on the statistics of curvature perturbations far beyond the range directly accessible to the cosmic microwave background. The collapse fraction is tiny for ordinary Gaussian amplitudes and grows rapidly once the tail of the distribution broadens. This exponential sensitivity has made primordial black holes a standard probe of rare-event statistics in inflationary cosmology \cite{Carr1974,Green1997,Carr2020,Green2021}.

The curvaton scenario occupies a special place in this subject. A light spectator field acquires nearly Gaussian field fluctuations during inflation. Those fluctuations remain subdominant during inflation, evolve into a distinct fluid after inflation, and later imprint the final curvature perturbation when the curvaton decays. The resulting map from the Gaussian field perturbation to the final curvature perturbation is intrinsically nonlinear. For sufficiently small curvaton fraction at decay, the probability density of the final curvature perturbation acquires a strongly distorted high-amplitude tail. The primordial black-hole abundance then becomes controlled by the exact nonlinear map instead of a low-order non-Gaussian truncation \cite{LythWands2002,Moroi2001,Sasaki2006,Young2013,Byrnes2012}.

The central goal of this paper is to connect exact curvaton non-Gaussianity to primordial black-hole phenomenology without mixing incompatible approximations. The exact curvaton map is derived once and then used consistently in the full probability density, in the abundance calculation, in the matched-variance comparisons, and in the scale-dependent mass-function forecasts. The scale-dependent section no longer fixes the curvaton non-Gaussian parameter independently at each smoothing scale. Instead, a single dimensionless curvaton field spectrum is specified, the smoothed field variance is computed directly, and the exact collapse fraction follows from the same decay fraction and the same field spectrum on every scale. The induced gravitational-wave section computes the standard radiation-era background from the linear curvaton two-point spectrum implied by the same self-consistent model. Additional genuinely non-Gaussian corrections to the induced gravitational-wave source are separate effects and remain outside the present scope.

The paper is organized as follows. Section 2 sets up the collapse problem in terms of the smoothed curvature perturbation, derives the Gaussian benchmark, and records the relation between formation fraction and present-day abundance. Section 3 derives the exact curvaton probability density from the sudden-decay relation, identifies the support of the distribution, and gives the exact primordial black-hole fraction. Section 4 introduces the exact local quadratic benchmark and explains the numerical pipeline for the benchmark figures. Section 5 presents the matched-variance benchmark results. Section 6 introduces the self-consistent scale-dependent curvaton model, derives the resulting mass functions, and confronts them with a conservative current primordial-black-hole constraint envelope. Section 7 derives the associated induced gravitational-wave spectra from the same model and displays them against the PTA, LISA, and DECIGO sensitivity windows. Section 8 discusses the physical interpretation and the assumptions of the calculation, and Section 9 summarizes the main conclusions.

\section{Primordial-black-hole formation from smoothed curvature perturbations}

We work with the curvature perturbation $\zeta_R$ smoothed on a comoving scale $R$. The primordial black-hole collapse problem depends on the probability that the smoothed perturbation exceeds a threshold of order unity when the corresponding scale re-enters the Hubble radius during radiation domination. In the present manuscript we express the collapse criterion directly in terms of $\zeta_R$ and write the formation fraction as
\begin{equation}
\beta(R) = \int_{\zeta_c}^{\infty} P_R(\zeta)\, d\zeta,
\label{eq:beta_def}
\end{equation}
where $P_R(\zeta)$ is the one-point probability density function of the smoothed curvature perturbation and $\zeta_c$ is an effective collapse threshold. Numerical simulations find threshold values in the broad interval $\zeta_c \sim 0.7$ to $1.2$, with the precise number controlled by the profile shape and the collapse variable \cite{ShibataSasaki1999,Harada2013,Harada2015,Musco2021}. The figures below use the representative choice $\zeta_c=1$, which is adequate for demonstrating the tail dependence and the internal hierarchy among the different non-Gaussian models.

The variance of the smoothed curvature perturbation is determined by the dimensionless scalar power spectrum $\mathcal{P}_\zeta(k)$ through
\begin{equation}
\sigma_\zeta^2(R) = \int_0^{\infty} d\ln k\, W^2(kR)\, \mathcal{P}_\zeta(k),
\label{eq:sigmaR}
\end{equation}
where $W(kR)$ is the smoothing window. In the numerical sections we use a Gaussian window,
\begin{equation}
W(kR)=\exp\!\left(-\frac{k^2R^2}{2}\right),
\label{eq:gaussian_window}
\end{equation}
because it yields stable one-dimensional integrals and keeps the scale assignment transparent.

For a Gaussian curvature perturbation the probability density is
\begin{equation}
P_R^{\rm G}(\zeta)=\frac{1}{\sqrt{2\pi}\,\sigma_\zeta(R)}\exp\!\left[-\frac{\zeta^2}{2\sigma_\zeta^2(R)}\right].
\label{eq:gaussian_pdf}
\end{equation}
Substituting Eq.~\eqref{eq:gaussian_pdf} into Eq.~\eqref{eq:beta_def} gives
\begin{equation}
\beta_{\rm G}(R)=\frac{1}{2}\,\mathrm{erfc}\!\left(\frac{\zeta_c}{\sqrt{2}\,\sigma_\zeta(R)}\right).
\label{eq:betaG_exact}
\end{equation}
The rare-event limit follows by using the large-argument asymptotic form of the complementary error function. Writing
\begin{equation}
\mathrm{erfc}(x)\simeq \frac{e^{-x^2}}{\sqrt{\pi}\,x}, \qquad x\gg 1,
\end{equation}
we obtain
\begin{equation}
\beta_{\rm G}(R)\simeq \frac{\sigma_\zeta(R)}{\zeta_c\sqrt{2\pi}}\exp\!\left[-\frac{\zeta_c^2}{2\sigma_\zeta^2(R)}\right].
\label{eq:betaG_asym}
\end{equation}
Equation \eqref{eq:betaG_asym} makes the main physical point explicit: the collapse fraction is exponentially sensitive to the ratio $\zeta_c/\sigma_\zeta$.

The mapping between scale and horizon mass during radiation domination is written in the standard form
\begin{equation}
M(k) \simeq 30\,M_\odot\left(\frac{\gamma}{0.2}\right)
\left(\frac{g_{\ast,f}}{10.75}\right)^{-1/6}
\left(\frac{k}{2.9\times 10^5\,\mathrm{Mpc}^{-1}}\right)^{-2},
\label{eq:Mk}
\end{equation}
where $\gamma$ is the collapse-efficiency factor and $g_{\ast,f}$ is the effective number of relativistic degrees of freedom at formation \cite{Carr2020,Green2021}. The present-day abundance associated with a narrow mass interval then scales as
\begin{equation}
\frac{d f_{\rm PBH}}{d\ln M}
\simeq
1.68\times 10^8
\left(\frac{\gamma}{0.2}\right)^{1/2}
\left(\frac{g_{\ast,f}}{106.75}\right)^{-1/4}
\left(\frac{M}{M_\odot}\right)^{-1/2}
\beta(M),
\label{eq:fPBHbeta}
\end{equation}
which follows from the fact that the primordial black-hole energy density scales as matter from formation onward, whereas the radiation bath redshifts as $a^{-4}$ until matter-radiation equality \cite{Carr2020}. The coefficient in Eq.~\eqref{eq:fPBHbeta} is the standard monochromatic approximation. In the extended-spectrum section we apply it scale by scale to display the relative and absolute size of the mass function.

\section{Exact curvaton non-Gaussianity}

We now derive the exact probability density of the curvature perturbation in the sudden-decay curvaton scenario. The derivation is the central element of the manuscript, so every algebraic step is written explicitly.

Let the curvaton energy fraction at decay be
\begin{equation}
\Omega_\chi \equiv \frac{\rho_\chi}{\rho_r+\rho_\chi}\Bigg|_{\rm dec}.
\label{eq:Omega_def}
\end{equation}
In the sudden-decay approximation, energy conservation across the decay hypersurface gives
\begin{equation}
(1-\Omega_\chi)e^{4(\zeta_r-\zeta)}+\Omega_\chi e^{3(\zeta_\chi-\zeta)}=1.
\label{eq:sudden_decay}
\end{equation}
The radiation contribution is set to $\zeta_r=0$ throughout the present analysis. Multiplying Eq.~\eqref{eq:sudden_decay} by $e^{4\zeta}$ yields
\begin{equation}
(1-\Omega_\chi)+\Omega_\chi e^{\zeta+3\zeta_\chi}=e^{4\zeta}.
\label{eq:after_mult}
\end{equation}
Rearranging gives
\begin{equation}
\Omega_\chi e^{3\zeta_\chi}=e^{3\zeta}+(\Omega_\chi-1)e^{-\zeta}.
\label{eq:e3zchi_raw}
\end{equation}
It is convenient to define
\begin{equation}
Y(\zeta)\equiv \frac{1}{\Omega_\chi}\left[e^{3\zeta}+(\Omega_\chi-1)e^{-\zeta}\right],
\label{eq:Ydef}
\end{equation}
so that Eq.~\eqref{eq:e3zchi_raw} becomes
\begin{equation}
e^{3\zeta_\chi}=Y(\zeta).
\label{eq:e3zchiY}
\end{equation}

For a quadratic oscillating curvaton, the local curvaton curvature perturbation satisfies
\begin{equation}
e^{3\zeta_\chi}=\left(1+\frac{\delta\chi}{\bar\chi}\right)^2,
\label{eq:chi_map}
\end{equation}
where $\bar\chi$ is the homogeneous curvaton background at the onset of oscillations and $\delta\chi$ is the Gaussian field fluctuation. Combining Eqs.~\eqref{eq:e3zchiY} and \eqref{eq:chi_map} gives
\begin{equation}
1+\frac{\delta\chi}{\bar\chi}=s\sqrt{Y(\zeta)}, \qquad s=\pm 1.
\label{eq:branch_eq}
\end{equation}
Solving for the field fluctuation on each branch yields
\begin{equation}
\delta\chi_s(\zeta)=\bar\chi\left[s\sqrt{Y(\zeta)}-1\right], \qquad s=\pm 1.
\label{eq:deltachi_exact}
\end{equation}
This expression is the branch map used throughout the figures.

The support of the probability density follows from the condition $Y(\zeta)\ge 0$. Equation \eqref{eq:Ydef} can be rewritten as
\begin{equation}
Y(\zeta)=\frac{e^{-\zeta}}{\Omega_\chi}\left[e^{4\zeta}-(1-\Omega_\chi)\right].
\label{eq:Ysupport}
\end{equation}
The prefactor $e^{-\zeta}/\Omega_\chi$ is positive, so the support condition becomes
\begin{equation}
e^{4\zeta}\ge 1-\Omega_\chi.
\end{equation}
The probability density therefore begins at the lower endpoint
\begin{equation}
\zeta_{\min}=\frac{1}{4}\ln(1-\Omega_\chi), \qquad 0<\Omega_\chi<1.
\label{eq:zmin}
\end{equation}
The derivative of $Y$ is
\begin{equation}
Y'(\zeta)=\frac{1}{\Omega_\chi}\left[3e^{3\zeta}-(\Omega_\chi-1)e^{-\zeta}\right]
=
\frac{1}{\Omega_\chi}\left[3e^{3\zeta}+(1-\Omega_\chi)e^{-\zeta}\right],
\label{eq:Yprime}
\end{equation}
which is positive on the support. The map from $\zeta$ to $Y$ is therefore monotonic.

Differentiating Eq.~\eqref{eq:deltachi_exact} with respect to $\zeta$ yields the branch Jacobian,
\begin{equation}
\frac{d\delta\chi_s}{d\zeta}
=
\bar\chi\,\frac{s\,Y'(\zeta)}{2\sqrt{Y(\zeta)}}.
\label{eq:jacobian}
\end{equation}
The underlying curvaton-field fluctuation is Gaussian,
\begin{equation}
P_\chi(\delta\chi)=\frac{1}{\sqrt{2\pi}\,\sigma_\chi}\exp\!\left[-\frac{\delta\chi^2}{2\sigma_\chi^2}\right],
\label{eq:Pchi}
\end{equation}
so the full curvature-perturbation probability density follows from a branchwise change of variables,
\begin{equation}
P_\zeta(\zeta)=\sum_{s=\pm} P_\chi\!\left(\delta\chi_s(\zeta)\right)
\left|\frac{d\delta\chi_s}{d\zeta}\right|.
\label{eq:Pz_general}
\end{equation}
Substituting Eqs.~\eqref{eq:deltachi_exact}, \eqref{eq:jacobian}, and \eqref{eq:Pchi} gives the explicit exact density,
\begin{equation}
P_\zeta(\zeta)=
\sum_{s=\pm}
\frac{1}{\sqrt{2\pi}\,\sigma_\chi}
\exp\!\left[-\frac{\bar\chi^2\left(s\sqrt{Y(\zeta)}-1\right)^2}{2\sigma_\chi^2}\right]
\left|\bar\chi\,\frac{s\,Y'(\zeta)}{2\sqrt{Y(\zeta)}}\right|.
\label{eq:Pz_explicit}
\end{equation}
The primordial black-hole formation fraction in the exact curvaton model is then
\begin{equation}
\beta_{\rm curv}(R;\Omega_\chi,\bar\chi/\sigma_\chi)=
\int_{\zeta_c}^{\infty} P_\zeta(\zeta)\, d\zeta.
\label{eq:beta_curv_exact}
\end{equation}

Only the ratio $\alpha\equiv \bar\chi/\sigma_\chi$ enters the exact density. Once $\Omega_\chi$ and $\alpha$ are fixed, the moments of the curvature perturbation are determined by
\begin{equation}
\langle \zeta^n\rangle = \int_{\zeta_{\min}}^{\infty} \zeta^n P_\zeta(\zeta)\, d\zeta.
\label{eq:moments}
\end{equation}
The matched-variance comparisons use $\Omega_\chi$ and the target variance $\sigma_\zeta^2$ as the control parameters. For each chosen $\Omega_\chi$ and each chosen $\sigma_\zeta$, the ratio $\alpha$ is determined by solving
\begin{equation}
\int_{\zeta_{\min}}^{\infty} \left(\zeta-\langle\zeta\rangle\right)^2 P_\zeta(\zeta)\, d\zeta
=
\sigma_\zeta^2.
\label{eq:variance_match}
\end{equation}
This calibration isolates the effect of the shape of the probability density from the trivial effect of changing the total variance.

The small-fluctuation expansion of the exact map is useful because it connects the exact treatment to the familiar local non-Gaussian coefficients. The derivation is given in Appendix A. Expanding Eq.~\eqref{eq:sudden_decay} to second order in $x\equiv\delta\chi/\bar\chi$ gives
\begin{equation}
\zeta = \frac{2\Omega_\chi}{4-\Omega_\chi}x
+ \frac{\Omega_\chi(16-8\Omega_\chi-\Omega_\chi^2)}{(4-\Omega_\chi)^3}x^2 + \mathcal{O}(x^3),
\label{eq:zeta_expand_main}
\end{equation}
and the induced local quadratic coefficient becomes
\begin{equation}
f_{\rm NL}=\frac{5}{4r_{\rm dec}}-\frac{5}{3}-\frac{5r_{\rm dec}}{6},
\qquad
r_{\rm dec}=\frac{3\Omega_\chi}{4-\Omega_\chi},
\label{eq:fNL_curv_main}
\end{equation}
which is the standard curvaton result \cite{Bartolo2004,Sasaki2006}. Appendix A shows the algebra explicitly.

\section{Exact local benchmark and numerical pipeline}

A matched-variance local quadratic benchmark is useful because it separates two distinct questions. The first question concerns the effect of a skewed nonlinear map at fixed variance. The second question concerns the extra structure carried by the exact curvaton map beyond a single local coefficient. The benchmark map is written as
\begin{equation}
\zeta = x + a\left(x^2-\sigma_x^2\right), \qquad a\equiv \frac{3}{5}f_{\rm NL},
\label{eq:local_quad_map}
\end{equation}
where $x$ is Gaussian with variance $\sigma_x^2$. Since the quadratic correction is centered, the mean of $\zeta$ is zero. Its variance is
\begin{equation}
\sigma_\zeta^2 = \langle \zeta^2\rangle
= \langle x^2\rangle + a^2\left\langle\left(x^2-\sigma_x^2\right)^2\right\rangle
= \sigma_x^2 + 2a^2\sigma_x^4.
\label{eq:local_var}
\end{equation}
For a target variance $\sigma_\zeta^2$, Eq.~\eqref{eq:local_var} fixes $\sigma_x^2$ through a quadratic equation,
\begin{equation}
2a^2\sigma_x^4 + \sigma_x^2 - \sigma_\zeta^2 = 0,
\label{eq:sigx_quad}
\end{equation}
so that
\begin{equation}
\sigma_x^2 = \frac{-1+\sqrt{1+8a^2\sigma_\zeta^2}}{4a^2}
\label{eq:sigx_sol}
\end{equation}
for $a\neq 0$.

To obtain the exact probability density generated by the benchmark map, we solve Eq.~\eqref{eq:local_quad_map} for the Gaussian variable $x$. Rearranging gives
\begin{equation}
a x^2 + x - a\sigma_x^2 - \zeta = 0.
\label{eq:local_quad_eq}
\end{equation}
The two roots are
\begin{equation}
x_\pm(\zeta)=\frac{-1\pm\sqrt{1+4a(\zeta+a\sigma_x^2)}}{2a}.
\label{eq:local_roots}
\end{equation}
The derivative of the map is
\begin{equation}
\frac{d\zeta}{dx}=1+2ax,
\label{eq:local_jac}
\end{equation}
so the exact local quadratic density is
\begin{equation}
P_\zeta^{\rm loc}(\zeta)=
\sum_{\pm}
\frac{1}{\sqrt{2\pi}\,\sigma_x}
\exp\!\left[-\frac{x_\pm^2(\zeta)}{2\sigma_x^2}\right]
\frac{1}{\left|1+2a x_\pm(\zeta)\right|}.
\label{eq:local_pdf_exact}
\end{equation}
The corresponding primordial black-hole fraction follows from the same threshold integral as before,
\begin{equation}
\beta_{\rm loc}(R)=\int_{\zeta_c}^{\infty} P_\zeta^{\rm loc}(\zeta)\, d\zeta.
\label{eq:beta_loc_exact}
\end{equation}
The local benchmark therefore enters the benchmark figures on the same footing as the exact curvaton density and the Gaussian density.

All figures were generated with a single internally consistent pipeline. The exact curvaton probability density in Eq.~\eqref{eq:Pz_explicit} was integrated over the support interval $[\zeta_{\min},\infty)$ and truncated numerically at $\zeta=8$, where the omitted probability mass was negligible for every parameter choice used in the paper. The numerical normalization error remained below $10^{-10}$ throughout the final figure set. For each matched-variance comparison, the parameter $\alpha=\bar\chi/\sigma_\chi$ was determined by solving Eq.~\eqref{eq:variance_match}. The exact collapse fraction then followed from Eq.~\eqref{eq:beta_curv_exact}. The local quadratic benchmark used Eqs.~\eqref{eq:sigx_sol} and \eqref{eq:local_pdf_exact}. The mass-function forecasts used Eqs.~\eqref{eq:sigmaR}, \eqref{eq:Mk}, and \eqref{eq:fPBHbeta}. The induced gravitational-wave spectra used the standard radiation-era kernel summarized in Appendix B together with the self-consistent two-point spectrum in Eq.~\eqref{eq:Pzeta_from_Px}.

\section{Corrected matched-variance results}

\begin{figure}[t]
\centering
\includegraphics[width=0.82\textwidth]{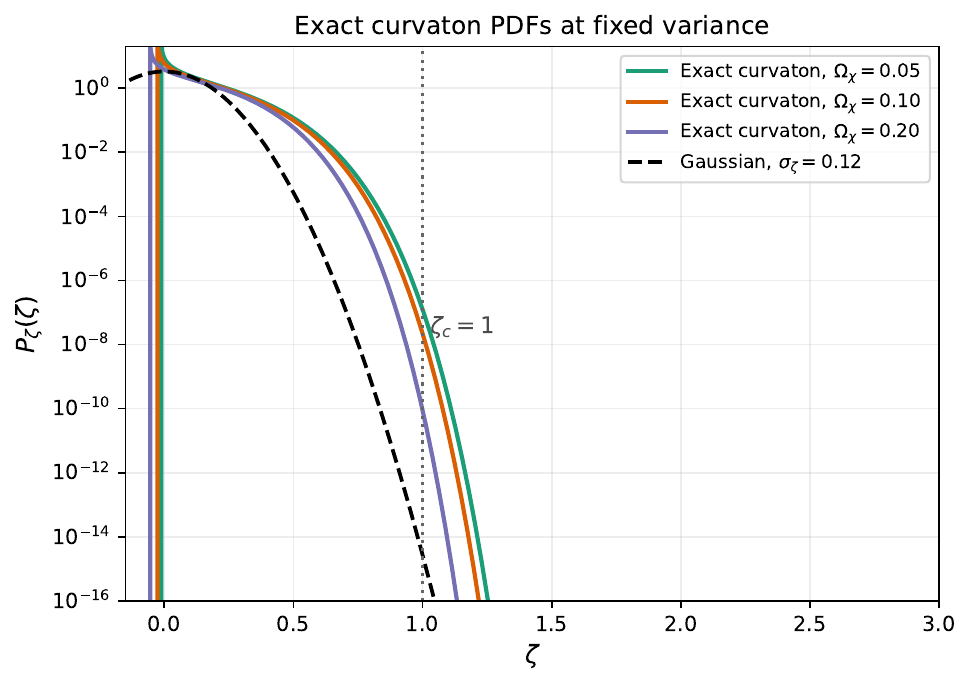}
\caption{Exact curvaton probability densities obtained from Eq.~\eqref{eq:Pz_explicit} after matching the variance to $\sigma_\zeta=0.12$. The dashed curve is the Gaussian benchmark with the same variance. The vertical dotted line marks the collapse threshold $\zeta_c=1$. Small curvaton fraction at decay produces a much broader high-amplitude tail.}
\label{fig:exact_pdfs}
\end{figure}

Figure~\ref{fig:exact_pdfs} shows the exact curvaton probability density for three decay fractions, with the variance fixed to $\sigma_\zeta=0.12$ in every case. The figure isolates the pure effect of the nonlinear tail shape. The smallest value of $\Omega_\chi$ produces the broadest upper tail and therefore the largest collapse fraction at fixed variance. Increasing $\Omega_\chi$ moves the exact density closer to the Gaussian benchmark and strongly reduces the probability weight above the collapse threshold.

\begin{figure}[t]
\centering
\includegraphics[width=0.82\textwidth]{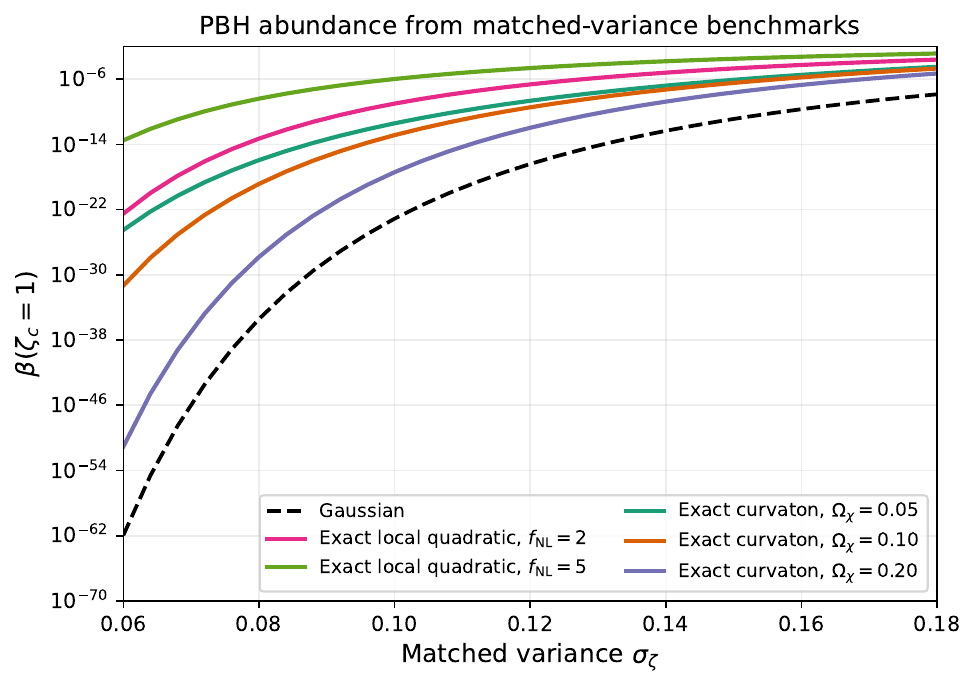}
\caption{Primordial black-hole fraction $\beta$ as a function of the matched variance $\sigma_\zeta$ for the Gaussian benchmark, two exact local quadratic benchmarks, and three exact curvaton models. Every curve uses the same collapse threshold $\zeta_c=1$. The exact curvaton enhancement is strongest for small curvaton decay fraction.}
\label{fig:beta_sigma}
\end{figure}

The dependence of the collapse fraction on the matched variance is displayed in Figure~\ref{fig:beta_sigma}. The Gaussian result follows the standard exponential suppression of Eq.~\eqref{eq:betaG_asym}. Positive local quadratic benchmarks lift the tail and generate a much larger $\beta$ at the same variance. The exact curvaton curves also produce strong enhancements for small $\Omega_\chi$. The figure therefore separates three levels of description. The Gaussian curve gives the reference scale, the local quadratic curves isolate the effect of a simple skewed nonlinear map, and the exact curvaton curves show how the full sudden-decay map reshapes the collapse tail.

\begin{figure}[t]
\centering
\includegraphics[width=0.82\textwidth]{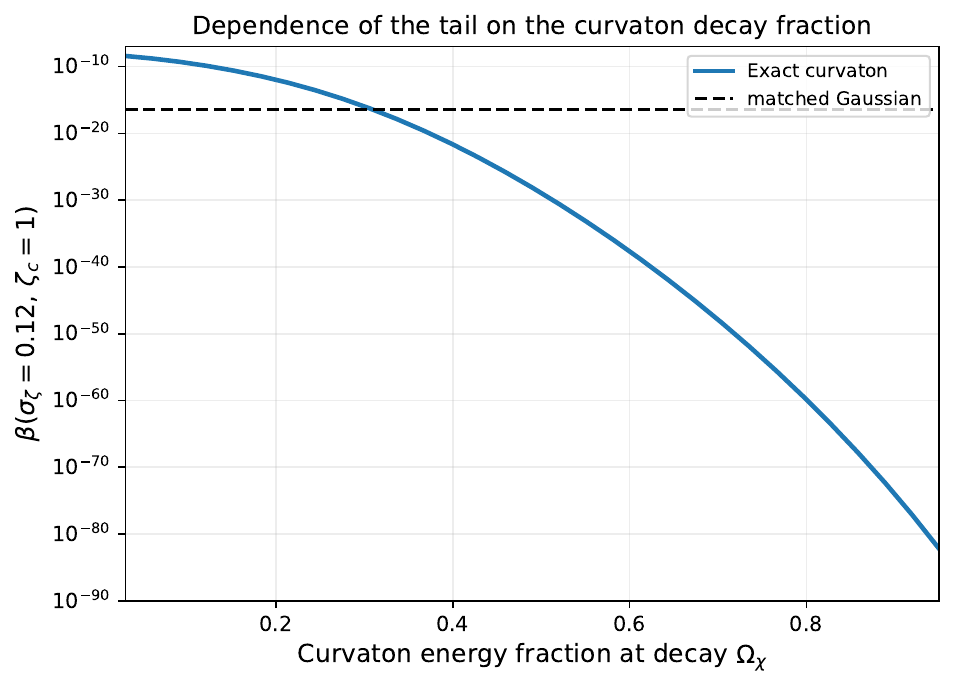}
\caption{Primordial black-hole fraction at fixed variance $\sigma_\zeta=0.12$ as a function of the curvaton decay fraction. The dashed horizontal line is the Gaussian reference at the same variance. The exact curvaton tail broadens rapidly as $\Omega_\chi$ decreases.}
\label{fig:beta_omega}
\end{figure}

Figure~\ref{fig:beta_omega} focuses on the curvaton decay fraction itself. The variance is again held fixed at $\sigma_\zeta=0.12$, so the horizontal scale measures a pure shape effect. The collapse fraction decreases monotonically as $\Omega_\chi$ grows. The change spans many orders of magnitude. This single figure summarizes the main reason that the decay fraction is a critical phenomenological control parameter: a modest change in the fluid composition at decay can move the probability density from a broad tail to a sharply suppressed tail even before any change in the total power is introduced.

Two remarks are important for the interpretation of Figures~\ref{fig:exact_pdfs}--\ref{fig:beta_omega}. First, the matched-variance comparison was designed to isolate the effect of the tail shape. It does not identify the unique microscopic relation between $\Omega_\chi$ and the power amplitude on every scale. Second, the exact curvaton density can produce either enhancement or suppression relative to a Gaussian of the same variance, depending on the region of parameter space. The figures shown here focus on the small-$\Omega_\chi$ domain because that is the regime most relevant for strong tail enhancement.

\section{Self-consistent scale-dependent curvaton model and observational confrontation}

The matched-variance figures isolate the pure tail effect, but they do not yet determine the scale dependence from a single curvaton fluctuation model. We now impose that additional structure explicitly. Let
\begin{equation}
x(\mathbf{x})\equiv \frac{\delta\chi(\mathbf{x})}{\bar\chi},
\end{equation}
so that the exact branch map in Eq.~\eqref{eq:deltachi_exact} depends on the smoothing scale only through the variance of the Gaussian variable $x$. We choose a single dimensionless curvaton fluctuation spectrum,
\begin{equation}
\mathcal{P}_x(k)=A_x\exp\!\left[-\frac{\ln^2(k/k_p)}{2\sigma_k^2}\right],
\label{eq:Px_model}
\end{equation}
with amplitude $A_x$, peak wavenumber $k_p$, and logarithmic width $\sigma_k$. Once Eq.~\eqref{eq:Px_model} is fixed, the smoothed variance of the Gaussian curvaton variable on scale $R$ follows directly from
\begin{equation}
\sigma_x^2(R)=\int_0^{\infty} d\ln q\, W^2(qR)\,\mathcal{P}_x(q).
\label{eq:sigxR}
\end{equation}
The exact probability density on that scale is then obtained by setting
\begin{equation}
\alpha(R)=\frac{\bar\chi}{\sigma_\chi(R)}=\frac{1}{\sigma_x(R)}.
\label{eq:alpha_of_R}
\end{equation}
No further fitting condition is introduced at each scale. The collapse fraction in the self-consistent curvaton model is therefore
\begin{equation}
\beta_{\rm curv}(R;\Omega_\chi,A_x,k_p,\sigma_k)
=
\int_{\zeta_c}^{\infty} P_\zeta\!\left(\zeta;\Omega_\chi,\alpha(R)\right)\,d\zeta,
\label{eq:beta_self_consistent}
\end{equation}
with $P_\zeta$ given by Eq.~\eqref{eq:Pz_explicit}. The present-day mass function follows by composing Eqs.~\eqref{eq:Mk}, \eqref{eq:fPBHbeta}, and \eqref{eq:beta_self_consistent}.

The same model also determines the two-point curvature spectrum that sources induced gravitational waves. From the small-fluctuation expansion in Appendix A, the linear transfer coefficient between the Gaussian curvaton fluctuation and the curvature perturbation is
\begin{equation}
A_1(\Omega_\chi)=\frac{2\Omega_\chi}{4-\Omega_\chi}.
\label{eq:A1transfer}
\end{equation}
Keeping the exact non-Gaussian map for the collapse problem while using the linear transfer for the two-point function, the curvature power spectrum entering the induced gravitational-wave kernel is
\begin{equation}
\mathcal{P}_\zeta^{(2)}(k)=A_1^2(\Omega_\chi)\,\mathcal{P}_x(k).
\label{eq:Pzeta_from_Px}
\end{equation}
Equation \eqref{eq:Pzeta_from_Px} is the precise sense in which the power spectrum and the exact tail are now tied to the same curvaton model.

To confront the scale-dependent mass functions with current primordial-black-hole limits without importing a large external data table into the manuscript, we use a conservative envelope $f_{\max}(M)$ motivated by the up-to-date review compilation of Carr et al. and by the open asteroid-mass window analysis of Montero-Camacho et al. \cite{CarrReview2026,MonteroCamacho2019}. The envelope is defined by log-log interpolation through the anchor points
\begin{equation}
\begin{aligned}
&(M_i/M_\odot,f_i)=\big(10^{-17},1\big),\ \big(10^{-12},1\big),\ \big(10^{-6},3\times 10^{-1}\big),\\
&\big(10^{-2},10^{-2}\big),\ \big(10^{-1},10^{-3}\big),\ \big(1,10^{-3}\big),\ \big(10^{2},10^{-3}\big),\ \big(10^{4},10^{-2}\big).
\end{aligned}
\label{eq:constraint_anchors}
\end{equation}
This envelope is intentionally conservative. It captures the broad current fact pattern most relevant for the present paper: the asteroid-mass window remains comparatively open, whereas the stellar-mass range is already strongly constrained. Appendix C writes the interpolation formula explicitly.

\begin{figure}[t]
\centering
\includegraphics[width=0.84\textwidth]{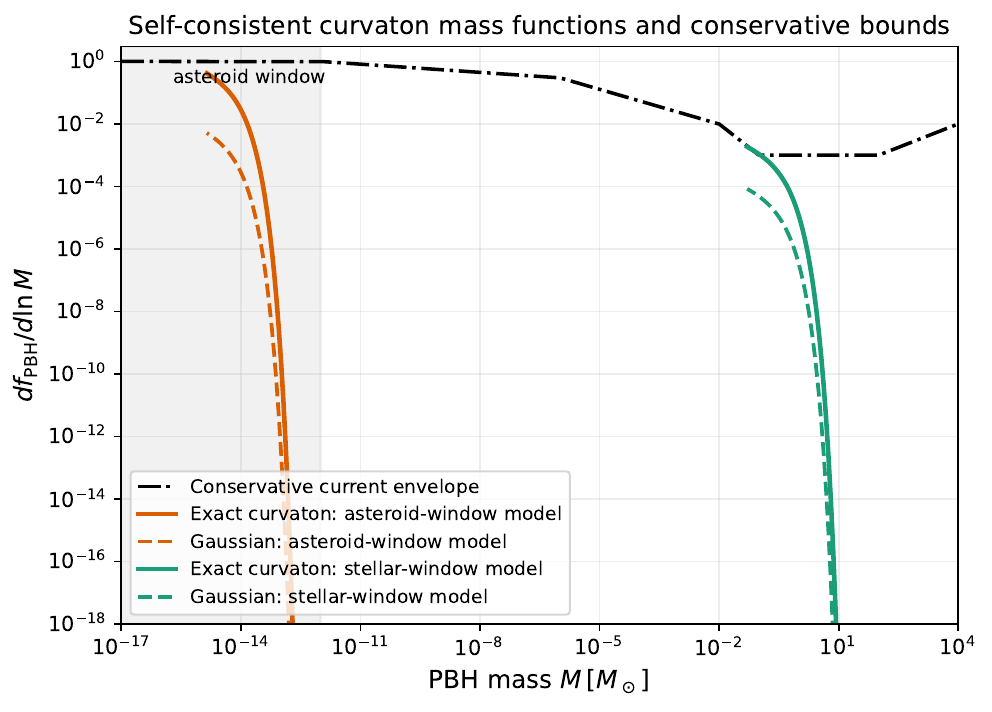}
\caption{Extended primordial black-hole mass functions obtained from the self-consistent curvaton fluctuation spectrum in Eq.~\eqref{eq:Px_model}. Solid curves use the exact curvaton probability density. Dashed curves use the Gaussian benchmark with the same variance implied by the same two-point spectrum. The black dot-dashed curve is the conservative current constraint envelope defined by Eq.~\eqref{eq:constraint_anchors}. The asteroid-window model remains allowed, whereas the stellar-window model is driven close to the current exclusion frontier.}
\label{fig:mass_constraints}
\end{figure}

Figure~\ref{fig:mass_constraints} shows two self-consistent curvaton models together with their Gaussian counterparts and with the conservative envelope in Eq.~\eqref{eq:constraint_anchors}. The first model uses $\Omega_\chi=0.20$, $A_x=0.95$, $k_p=3\times 10^{12}\,\mathrm{Mpc}^{-1}$, and $\sigma_k=0.50$. Its mass function peaks in the asteroid window and remains below the conservative envelope. The second model uses $\Omega_\chi=0.20$, $A_x=1.35$, $k_p=5\times 10^{5}\,\mathrm{Mpc}^{-1}$, and $\sigma_k=0.50$. Its mass function peaks near the stellar range and approaches the current envelope from below. In both cases the exact curvaton tail lifts the mass function far above the Gaussian prediction computed from the same two-point spectrum. The comparison therefore cleanly separates the role of the exact non-Gaussian tail from the role of the underlying scale dependence.

\section{Induced gravitational waves and detector windows}

We now use the same self-consistent curvaton models to compute the induced gravitational-wave background. The present-day spectrum is written as
\begin{equation}
\Omega_{\rm GW,0}(k)h^2=0.83\,\Omega_{r,0}h^2\,\Omega_{\rm GW,c}(k),
\label{eq:OmegaGWtoday}
\end{equation}
where $\Omega_{r,0}h^2\simeq 4.15\times 10^{-5}$ and $\Omega_{\rm GW,c}(k)$ is the oscillation-averaged radiation-era source spectrum. The radiation-era kernel is given in Appendix B. The only change is that the input scalar spectrum is no longer imposed directly at the curvature level. Instead, Eq.~\eqref{eq:Pzeta_from_Px} carries the curvaton field spectrum into the two-point curvature spectrum. This ties the frequency-domain signal to the same microscopic parameters that control the exact collapse tail.

\begin{figure}[t]
\centering
\includegraphics[width=0.84\textwidth]{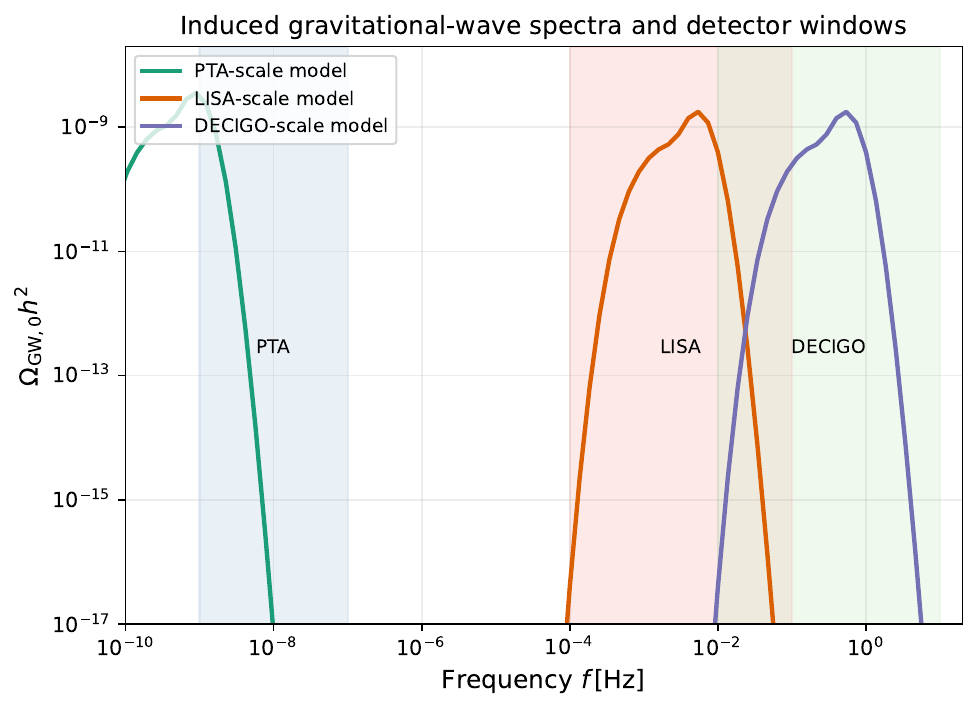}
\caption{Induced gravitational-wave spectra generated from the self-consistent curvaton field spectrum through Eq.~\eqref{eq:Pzeta_from_Px}. The shaded vertical bands mark the approximate PTA, LISA, and DECIGO sensitivity windows. Moving the peak wavenumber of the curvaton feature moves the induced signal across the detector landscape.}
\label{fig:gw_windows}
\end{figure}

Figure~\ref{fig:gw_windows} displays three representative spectra. The first uses the asteroid-window mass-function parameters and therefore peaks in the LISA band. The second uses the stellar-window mass-function parameters and therefore peaks in the PTA band. The third uses the same $\Omega_\chi$ and $A_x=0.95$ as the asteroid model but shifts the peak to $k_p=3\times 10^{14}\,\mathrm{Mpc}^{-1}$, which moves the signal into the decihertz range. The shaded regions mark the broad sensitivity windows associated with pulsar timing arrays, LISA, and DECIGO, using the standard nanohertz, millihertz, and decihertz bands emphasized in current detector summaries and mission studies \cite{NANOGrav2023,LISA2018,Seto2001}. The figure shows directly how changing the location of the curvaton feature moves the gravitational-wave signal across the detector landscape while the exact non-Gaussian tail continues to govern the primordial-black-hole abundance.

\section{Discussion and Conclusion}

The manuscript establishes a clean separation between two distinct physical effects. The first effect is the shape of the probability density at fixed variance. That effect is governed here by the exact sudden-decay curvaton map. The second effect is the scale dependence of the variance itself. In the final sections that second effect is no longer imposed at the curvature level by hand. It is generated from a single dimensionless curvaton field spectrum, so the same microscopic parameters determine the smoothed field variance, the exact collapse tail, the mass function, and the linear two-point spectrum entering the induced gravitational-wave kernel.

The exact curvaton probability density carries two branch contributions, both of which are retained in the numerical pipeline. The lower support boundary in Eq.~\eqref{eq:zmin} is imposed explicitly in every integral. In the benchmark section the variance is matched by solving for $\alpha=\bar\chi/\sigma_\chi$ at fixed $\sigma_\zeta$. In the scale-dependent section the procedure is tighter. Equation~\eqref{eq:Px_model} determines $\sigma_x(R)$ directly, Eq.~\eqref{eq:alpha_of_R} fixes $\alpha(R)$, and the exact collapse fraction then follows with no additional scale-by-scale fitting. The price is that the field spectrum is still modeled phenomenologically by a lognormal feature. A more microscopic inflationary construction would derive the same feature from the curvaton background dynamics and its effective mass during inflation.

The induced gravitational-wave calculation in Section 7 uses the standard Gaussian scalar source term for the second-order radiation-era kernel together with the self-consistent curvaton two-point spectrum in Eq.~\eqref{eq:Pzeta_from_Px}. That choice is deliberate. It produces a quantitatively controlled baseline for the frequency structure and the amplitude scaling with the curvaton feature. Additional non-Gaussian corrections to the induced source can be important when higher-point functions are large, and the exact curvaton scenario offers a natural arena for that problem. A dedicated treatment would require the unequal-time scalar four-point structure of the curvaton-sourced perturbations. The present paper therefore treats the exact non-Gaussianity in the collapse problem and the linear two-point spectrum in the induced gravitational-wave kernel, keeping the two sectors tied by the same curvaton model while leaving the higher-order source correction to future work.

The collapse threshold was fixed at $\zeta_c=1$ throughout the numerical figures. A more detailed phenomenological study could scan over the threshold, include profile dependence through the compaction function, and incorporate critical-collapse broadening. The observational confrontation was also kept conservative: instead of importing a fully digitized exclusion table for extended mass functions, we used the envelope in Eq.~\eqref{eq:constraint_anchors} so that the comparison remains transparent in analytic form. Those refinements would change the numerical normalization of the mass functions and would sharpen the boundary between allowed and excluded models. The central conclusion would remain the same because it is driven by the far-tail sensitivity of the exact probability density and by the direct mapping between feature scale and gravitational-wave frequency.

This paper has carried the curvaton primordial-black-hole calculation from the exact sudden-decay relation to a scale-dependent phenomenology built from a single curvaton fluctuation model. The exact branch map from the Gaussian curvaton fluctuation to the final curvature perturbation has been derived explicitly. The support of the exact probability density, the Jacobian, the normalization, the threshold integral, and the small-fluctuation expansion have all been written in closed form. The benchmark figures isolate the pure effect of the non-Gaussian tail at fixed variance. The final phenomenological sections then replace the scale-by-scale variance-matching ansatz by a self-consistent dimensionless curvaton field spectrum, compute the smoothed field variance directly, and generate the exact collapse fraction, the present-day mass function, and the linear two-point spectrum from the same parameter set.

The results show that the curvaton decay fraction is a sharp control parameter for the primordial-black-hole tail. At fixed variance, small $\Omega_\chi$ broadens the upper tail and can enhance the formation fraction by many orders of magnitude. Once the scale dependence is derived from a single curvaton field spectrum, the exact tail continues to control the normalization of the mass function, while the peak scale controls the mass at which the signal appears. The confrontation with the conservative current constraint envelope shows that the same formalism can accommodate an asteroid-window model that remains allowed and a stellar-window model that is already pushed close to exclusion. The induced gravitational-wave spectra generated from the same self-consistent curvaton model move across the PTA, LISA, and DECIGO bands as the feature scale is shifted, which exposes the complementary role of gravitational-wave windows and primordial-black-hole bounds.

The next theoretical step is to promote the phenomenological curvaton field spectrum in Eq.~\eqref{eq:Px_model} to a fully dynamical inflationary construction and to compute the genuinely non-Gaussian correction to the induced gravitational-wave source itself. The present analysis provides the explicit mathematical foundation and the corrected numerical baseline for that program.

\clearpage
\appendix

\section{Small-fluctuation expansion of the exact curvaton map}

This appendix derives the local quadratic expansion quoted in Eqs.~\eqref{eq:zeta_expand_main} and \eqref{eq:fNL_curv_main}. Let
\begin{equation}
x\equiv \frac{\delta\chi}{\bar\chi}.
\end{equation}
Equation \eqref{eq:chi_map} gives
\begin{equation}
e^{3\zeta_\chi}=(1+x)^2=1+2x+x^2.
\label{eq:e3zchi_expand}
\end{equation}
Equation \eqref{eq:e3zchi_raw} can be written as
\begin{equation}
\Omega_\chi e^{3\zeta_\chi}=e^{3\zeta}+(\Omega_\chi-1)e^{-\zeta}.
\label{eq:appendix_master}
\end{equation}
Expanding the right-hand side in powers of $\zeta$ gives
\begin{align}
e^{3\zeta} &= 1+3\zeta+\frac{9}{2}\zeta^2+\mathcal{O}(\zeta^3),\\
(\Omega_\chi-1)e^{-\zeta} &= (\Omega_\chi-1)\left(1-\zeta+\frac{1}{2}\zeta^2+\mathcal{O}(\zeta^3)\right).
\end{align}
Adding the two series yields
\begin{equation}
e^{3\zeta}+(\Omega_\chi-1)e^{-\zeta}
=
\Omega_\chi + (4-\Omega_\chi)\zeta + \frac{8+\Omega_\chi}{2}\zeta^2 + \mathcal{O}(\zeta^3).
\label{eq:RHS_series}
\end{equation}
The left-hand side of Eq.~\eqref{eq:appendix_master} becomes
\begin{equation}
\Omega_\chi e^{3\zeta_\chi} = \Omega_\chi(1+2x+x^2).
\label{eq:LHS_series}
\end{equation}
We now solve for $\zeta$ as a power series in $x$,
\begin{equation}
\zeta = A x + B x^2 + \mathcal{O}(x^3).
\label{eq:zeta_ansatz}
\end{equation}
Substituting Eq.~\eqref{eq:zeta_ansatz} into Eq.~\eqref{eq:RHS_series} gives
\begin{equation}
\Omega_\chi + (4-\Omega_\chi)(Ax+Bx^2) + \frac{8+\Omega_\chi}{2}A^2x^2 + \mathcal{O}(x^3).
\label{eq:RHS_xseries}
\end{equation}
Matching the coefficient of $x$ between Eqs.~\eqref{eq:LHS_series} and \eqref{eq:RHS_xseries} yields
\begin{equation}
2\Omega_\chi = (4-\Omega_\chi)A,
\end{equation}
so that
\begin{equation}
A = \frac{2\Omega_\chi}{4-\Omega_\chi}.
\label{eq:Acoef}
\end{equation}
Matching the coefficient of $x^2$ gives
\begin{equation}
\Omega_\chi = (4-\Omega_\chi)B + \frac{8+\Omega_\chi}{2}A^2,
\end{equation}
which implies
\begin{equation}
B = \frac{\Omega_\chi - \frac{8+\Omega_\chi}{2}A^2}{4-\Omega_\chi}
=
\frac{\Omega_\chi(16-8\Omega_\chi-\Omega_\chi^2)}{(4-\Omega_\chi)^3}.
\label{eq:Bcoef}
\end{equation}
This reproduces Eq.~\eqref{eq:zeta_expand_main}.

To express the result in the usual local form, define the Gaussian linear piece
\begin{equation}
\zeta_g \equiv A x.
\end{equation}
Then Eq.~\eqref{eq:zeta_ansatz} becomes
\begin{equation}
\zeta = \zeta_g + \frac{B}{A^2}\zeta_g^2 + \mathcal{O}(\zeta_g^3).
\label{eq:local_compare}
\end{equation}
Comparing with
\begin{equation}
\zeta = \zeta_g + \frac{3}{5}f_{\rm NL}\left(\zeta_g^2-\langle \zeta_g^2\rangle\right)+\cdots,
\end{equation}
we identify
\begin{equation}
f_{\rm NL}=\frac{5}{3}\frac{B}{A^2}
=\frac{5(\Omega_\chi^2+24\Omega_\chi-16)}{12\Omega_\chi(\Omega_\chi-4)}.
\label{eq:fNL_Omega}
\end{equation}
Using
\begin{equation}
r_{\rm dec}=\frac{3\rho_\chi}{4\rho_r+3\rho_\chi}\Bigg|_{\rm dec}=\frac{3\Omega_\chi}{4-\Omega_\chi},
\end{equation}
Eq.~\eqref{eq:fNL_Omega} becomes
\begin{equation}
f_{\rm NL}=\frac{5}{4r_{\rm dec}}-\frac{5}{3}-\frac{5r_{\rm dec}}{6}.
\end{equation}
This is the familiar curvaton local quadratic coefficient obtained directly from the exact sudden-decay map.

\section{Radiation-era induced gravitational-wave kernel}

The induced gravitational-wave calculation used in Section 6 follows the standard radiation-era formalism \cite{Ananda2007,Baumann2007,KohriTerada2018}. The oscillation-averaged spectrum at late times during radiation domination is
\begin{equation}
\Omega_{\rm GW,c}(k)
=
\frac{1}{6}
\int_0^{\infty} dv
\int_{|1-v|}^{1+v} du\,
\left[\frac{4v^2-(1+v^2-u^2)^2}{4uv}\right]^2
\mathcal{P}_\zeta(kv)\mathcal{P}_\zeta(ku)
\overline{I^2_{\rm RD}(u,v)},
\label{eq:OmegaGWkernel}
\end{equation}
with the radiation-era kernel
\begin{equation}
\begin{aligned}
\overline{I^2_{\rm RD}(u,v)}=
&\frac{1}{2}\left[\frac{3(u^2+v^2-3)}{4u^3v^3}\right]^2 \\
&\times\Bigg\{
\left[-4uv+(u^2+v^2-3)\ln\left|\frac{3-(u+v)^2}{3-(u-v)^2}\right|\right]^2 \\
&\hspace{1.2cm}+\pi^2(u^2+v^2-3)^2\Theta(u+v-\sqrt{3})
\Bigg\}.
\end{aligned}
\label{eq:Ibar}
\end{equation}
Substituting Eq.~\eqref{eq:OmegaGWkernel} into Eq.~\eqref{eq:OmegaGWtoday} gives the present-day spectrum. The numerical spectra in Figure~\ref{fig:gw_windows} were obtained by direct evaluation of this kernel after replacing the scalar power spectrum by the self-consistent curvaton two-point spectrum in Eq.~\eqref{eq:Pzeta_from_Px}.

\section{Conservative constraint envelope for the observational comparison}

The observational comparison in Section 6 uses an analytic envelope rather than a digitized exclusion table. Let
\begin{equation}
\mu\equiv \ln\!\left(\frac{M}{M_\odot}\right),\qquad y(\mu)\equiv \ln f_{\max}(M).
\end{equation}
Define the anchor set $\{(\mu_i,y_i)\}$ by taking the logarithm of the mass-fraction pairs in Eq.~\eqref{eq:constraint_anchors}. On each interval $\mu_i\le \mu\le \mu_{i+1}$ we set
\begin{equation}
y(\mu)=y_i+\frac{y_{i+1}-y_i}{\mu_{i+1}-\mu_i}(\mu-\mu_i),
\end{equation}
which is linear interpolation in the $(\ln M,\ln f)$ plane. Exponentiating gives
\begin{equation}
f_{\max}(M)=\exp\!\big(y(\mu)\big).
\end{equation}
This construction is transparent, conservative, and sufficient for the qualitative confrontation carried out in Figure~\ref{fig:mass_constraints}. It should be replaced by a full digitized extended-mass-function likelihood analysis in any precision follow-up study.

\end{document}